**Title: The relationships between $PM_{2.5}$ and meteorological factors in China: Seasonal and regional variations**

**Authors:** Qianqian Yang [a], Qiangqiang Yuan[*] [a], Tongwen Li [b], Huanfeng Shen [b], Liangpei Zhang [c]

**Affiliations:**

[a] School of Geodesy and Geomatics, Wuhan University, Wuhan, Hubei, 430079, China.

[b] School of Resource and Environmental Sciences, Wuhan University, Wuhan, Hubei, 430079, China

[c] State Key Laboratory of Information Engineering, Survey Mapping and Remote Sensing, Wuhan University, Wuhan, Hubei, 430079, China

[*] **Corresponding author:**

Qiangqiang Yuan (yqiang86@gmail.com)

**ABSTRACT**

The interactions between $PM_{2.5}$ and meteorological factors play a crucial role in air pollution analysis. However, previous studies that have researched the relationships between $PM_{2.5}$ concentration and meteorological conditions have been mainly confined to a certain city or district, and the correlation over the whole of China remains unclear. Whether or not spatial and seasonal variations exit deserves further research. In this study, the relationships between $PM_{2.5}$ concentration and meteorological factors were investigated in 74 major cities in China for a continuous period of 22 months from February 2013 to November 2014, at season, year, city, and regional scales, and the spatial and seasonal variations were analyzed. The meteorological factors were relative humidity (RH), temperature (TEM), wind speed (WS), and surface pressure (PS). We found that spatial and seasonal variations of their relationships with $PM_{2.5}$ do exist. Spatially, RH is positively correlated with $PM_{2.5}$ concentration in North China and Urumqi, but the relationship turns to negative in other areas of China. WS is negatively correlated with $PM_{2.5}$ everywhere expect for Hainan Island. PS has a strong positive relationship with $PM_{2.5}$ concentration in Northeast China and Mid-south China, and in other areas the correlation is weak. Seasonally, the positive correlation between $PM_{2.5}$ concentration and RH is stronger in winter and spring. TEM has a negative relationship with $PM_{2.5}$ in autumn and the opposite in winter. PS is more positively correlated with $PM_{2.5}$ in autumn than in other seasons. Our study investigated the relationships between $PM_{2.5}$ and meteorological factors in terms of spatial and seasonal variations, and the conclusions about the relationships between $PM_{2.5}$ and meteorological factors are more comprehensive and precise than before. We suggest that the variations could be considered in $PM_{2.5}$ concentration prediction and haze control to improve the prediction accuracy and policy efficiency.

## 1. Introduction

Clean air is a basic requirement of human comfort, health, and well-being (Luo et al., 2017). However, in recent decades, with the rapid development of the economy, air pollution has grown more and more serious in China (Li et al., 2016). A study by the World Health Organization (WHO) showed that the premature deaths of more than two million people each year can be attributed to the effects of air pollution during the 21st century, and more than half of these deaths happened in developing countries, particularly China and India (World Health Organization, 2005). The serious adverse effect of air pollution makes it a hotspot of public concern and scientific study. $PM_{2.5}$ (particulate matter with an aerodynamic diameter of less than 2.5 $\mu$m) is one of the most harmful components in air pollutants (Zhang et al., 2014). A high concentration of $PM_{2.5}$ can have an adverse impact on human health, climate change, local ecosystems, and economic development (Wu et al., 2017; Guo et al., 2016; Lu et al., 2017; Zhao et al., 2006; Liu et al., 2017; Xie et al., 2016; Cao et al, 2016). As former studies have showed, high $PM_{2.5}$ concentrations can be attributed not only to increasing emissions (Dai et al., 2013; Lai et al., 2016), but also to many natural geographical factors, such as topography, vegetation, and climate (Luo et al., 2017, Cai et al., 2017, Liu et al., 2017; Zhang et al., 2017; Chang et al., 2017). Among these influencing factors, meteorological conditions are some of the most important factors (Jones et al., 2010). The study of the relationships between meteorological conditions and $PM_{2.5}$ concentration could help us to have a better understanding of the $PM_{2.5}$ pollution problem, and could contribute to the adoption of more effective measures to reduce $PM_{2.5}$ pollution.

To date, there have been many studies about the relationships between $PM_{2.5}$ concentration and meteorological conditions. For example, Chen et al. (2016) studied the relationships between $PM_{2.5}$ and meteorological factors in the Nanjing urban area from 2013 to 2015, and found a negative correlation between $PM_{2.5}$ concentration and wind speed, temperature, relative humidity, and precipitation. Li et al. (2015) conducted a similar analysis in the Sichuan Basin, and found that $PM_{2.5}$ concentration was negatively correlated with wind speed and air temperature, positively correlated with air pressure, and weakly correlated with relative humidity. The relationships between $PM_{2.5}$ concentration and meteorological factors have also been examined in many other cities such as Beijing, Shanghai, Guangzhou, and Wuhan, and areas such as Beijing-Tianjin-Hebei (BTH) (Huang et al., 2015; Xu et al., 2015; Zhang et al., 2015; Zhang et al., 2017; Chen et al., 2017). Many studies have already proved that the correlations between $PM_{2.5}$ concentration and meteorological factors vary with the seasons. For example, Chen et al. (2016) found that the correlation between $PM_{2.5}$ and TEM was negative in summer and autumn and then turned to positive in spring and winter. Chen et al. (2017) found that wind was negatively correlated with $PM_{2.5}$ in winter, but the correlation was weak in summer. However, besides seasonal variations, it has also been found that the relationships between $PM_{2.5}$ concentration and meteorological factors vary between regions. Taking the correlation between $PM_{2.5}$ concentration and relative humidity as an example, Huang et al. (2015) found a positive correlation between $PM_{2.5}$ concentration and relative humidity in Beijing, but Chen et al. (2016) reported a negative correlation in the Nanjing urban area. As for Li et al. (2015), they came to the conclusion that $PM_{2.5}$ concentration was weakly correlated with relative humidity in the Sichuan Basin. $PM_{2.5}$ concentration was also found to be negatively correlated with WS in Guangzhou in a study by Zhang et al. (2015), but in another analysis in Shijiazhuang, the result was a weak correlation (Chen et al., 2017). The reason for the spatial variations is that meteorological factors affect $PM_{2.5}$ concentration through a series of conflicting processes, and the final effect is a synthetic result depending on the local climate, terrain, and $PM_{2.5}$ components. For example, on the one hand, high temperature can promote the formation of secondary aerosols (Li et at., 2013) and increase $PM_{2.5}$ concentration but, on the other hand, high temperature can also promote the convection of air (Luo et al., 2017), thus creating better diffusion conditions for particulate matter and decreasing $PM_{2.5}$ concentration. Temperature can also influence $PM_{2.5}$ concentration through promoting the volatilization of ammonium nitrate (Tai et al., 2010) and affect the emission rates from domestic heating and power production (Tran et al., 2011). In China, the sources and components of $PM_{2.5}$, the local climate, and terrain are very complicated (Wang et al., 2014; Daly et al., 2002), making the influencing degree of the different processes vary between regions. As a result, the relationships between $PM_{2.5}$ concentration and meteorological factors are spatially heterogeneous. Therefore, only through comparisons between different regions can we describe the relationships between $PM_{2.5}$ concentration and meteorological factors more precisely and in more detail. Previous studies that were mainly confined to one city or district were not able to comprehensively investigate the relationships between $PM_{2.5}$ and meteorological factors. Correspondingly, their conclusions can only be applied to local regions and cannot provide useful suggestions for governmental decisions at a national scale. There have been some studies at a national scale (Luo et al., 2017), but in their analysis, China was regarded as one district, and the variations between the different regions were still ignored. At all events, the neglect of the regional variations makes the relationships between $PM_{2.5}$ concentration and meteorological factors still unclear in China. What is required is analyses including a larger number of cities and areas, and the variations among different regions and seasons need to be examined and summarized.

The goal of our study was to comprehensively investigate the relationships between $PM_{2.5}$ concentration and meteorological factors, so as to obtain a clear picture of their relationships in China. In this study, we conducted a correlation analysis in 74 major cities and seven geographic regions in China based on a 22-month record of observations for 2013–2014. The correlation coefficient (r) values of the different cities and areas were then compared and the varying patterns summarized. The seasonal differences of the relationships were also analyzed. The conclusions of this study could be utilized to improve the understanding of the formation mechanisms of air pollution and the accuracy of $PM_{2.5}$ forecasting, and could provide a reference for environmental management policy decision-making.

The rest of this paper is organized as follows. Section 2 is the data preparation part, where we introduce the study area and period, the data sources, the data preprocessing work, and the methodology, and provide a flow chart of the study design. The experimental results and a discussion are provided in Section 3. Finally, we make a summary of our work and discuss the future work in Section 4.

## 2. Data and method

*2.1. Study area and period*

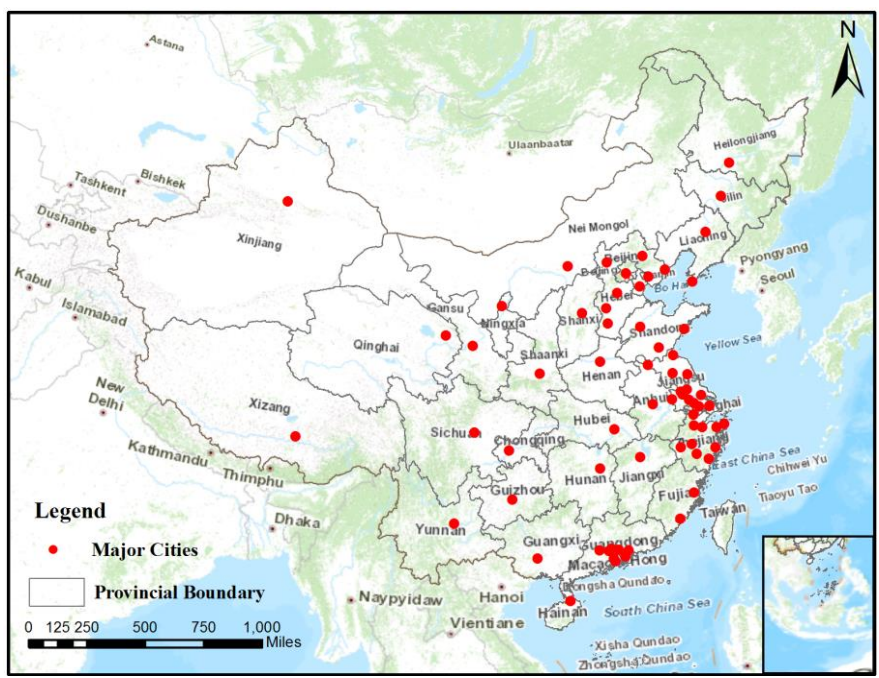

**Fig. 1.** Study area: 68 major cities in China.

China is the third largest country in the world, and has the largest population of any country. In recent years, it has witnessed a great leap in the economy; however, the rapid growth of Chinese cities has also resulted in lots of pollution, and China has already become one of the most seriously polluted countries in the world. Our study concentrated on the 74 major cities in China, as defined by the China National Ambient Air Quality Standard (CNAAQS, GB3095-2012), including the BTH, Yangtze River Delta (YRD), and Pearl River Delta (PRD) regions and municipalities, capital cities of the first-level administrative divisions, and other cities listed in the State Economic Plan. Because of insufficient data, six cities (Handan, Langfang, Cangzhou, Hengshui, Jiaxing, Jiangmen) were not included in the final correlation analysis. The locations of the other 68 cities are shown in Fig. 1. Although most of the cities are located in the PRD, YRD, and BTH regions, the 68 cities are spread over all of China, making the analysis of regional variations feasible. The time range of our study was from February 3, 2013, to November 30, 2014, i.e., 22 months/666 days in total. Due to the instrument measurement error, data transmission error, and other reasons, data of some days were missing, and thus the number of days of data actually used in the calculation was slightly less than 666 days. Four meteorological factors were studied in our research: relative humidity (RH), temperature (TEM), wind speed (WS), and surface pressure (PS).

*2.2. Data collection*

Since 2012, the 74 major cities in China have started building $PM_{2.5}$ monitoring sites, and air quality data have been published online since January 2013. We downloaded the daily $PM_{2.5}$ concentration data from the Data Center of the Ministry of Environmental Protection of the People's Republic of China (http://datacenter.mep.gov.cn/index), from 2013 to 2015. By December 2015, the number of $PM_{2.5}$ monitoring station exceeded 1440. The distribution of these monitoring sites in 2015 is shown in Fig. 2(a).

Meteorological data were downloaded from the China Meteorological Data Network (http://data.cma.cn/site/index.html). Relative humidity data were obtained from China's Ground Daily Climatological Data Collection, with the temporal resolution of one day and the time system of Beijing Time. Temperature, wind speed, and air pressure data were obtained from the Global Ground Meteorological Real-time Data Collection, with a temporal resolution of 3 hours and the time system of Greenwich Mean Time. There are numerous meteorological sites in China, and most of them are distributed evenly, as shown in Fig. 2(b).

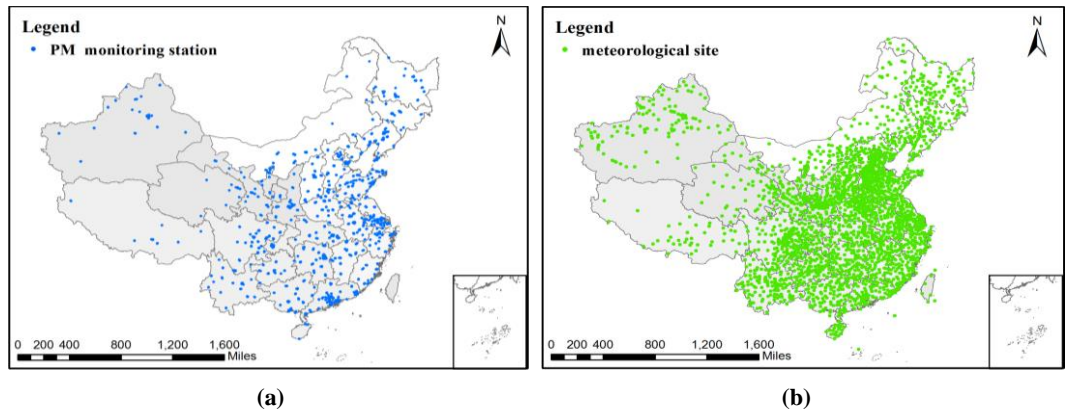

**Fig. 2.** The distribution of (a) PM$_{2.5}$ monitoring stations and (b) meteorological stations in China in 2015.

*2.3. Data preprocessing*

As shown in Fig. 2, PM$_{2.5}$ concentration and meteorological parameters are not measured at the same sites, so we needed to match the two types of data to form data pairs for the correlation calculation task. As the correlation analysis was conducted at the basic unit of city, we first screened the PM$_{2.5}$ monitoring sites for each city through setting longitude and latitude thresholds. We then set buffers centered on the PM$_{2.5}$ monitoring stations, and the buffer radius was determined to be 0.3° after a parameter sensitivity test. The parameter sensitivity results are shown in Supplementary Table 1. Meteorological sites located within the buffers were recorded. When the screening for one city was completed, we averaged the PM$_{2.5}$ concentration data and meteorological data of all the monitoring sites in this city area as the city-level data. There were a few outliers in the raw data which may have introduced errors, so a validity check on the daily data was conducted to remove the problematic data. We set a normal value threshold for each parameter, including PM$_{2.5}$ concentration, relative humidity, temperature, wind speed, and surface pressure, and the data falling outside of the thresholds were assigned as null. We then processed the data for correlation calculation at season, year, city, and regional scales.

*2.4. Methodology*

The process of our study design can be divided into four parts. The first step was the data preprocessing, as mentioned before. We then conducted a simple analysis of the spatiotemporal distribution pattern of $PM_{2.5}$ concentration. ArcGIS and the empirical Bayesian kriging (EBK) interpolation method (Krivoruchko et al., 2012) were used to acquire the spatial distribution map. Thirdly, we conducted the correlation analysis at multiple scales, i.e., city, region, season, and year. The correlations between daily $PM_{2.5}$ concentration and daily meteorological factors were measured by the Spearman's rank correlation coefficient. We first divided China into seven geographic regions and conducted the correlation analyses in both the 68 cities and seven geographic regions, to explore the spatial variations of the correlation at different spatial scales. The correlation analyses for the four seasons and whole year were then implemented. Finally, we analyzed the seasonal and regional variations of the relationships between $PM_{2.5}$ concentration and the meteorological factors. We drew the correlation coefficient (r) distribution maps of the 68 cities and seven regions in the four seasons and at a yearly scale, and conducted clustering to help with the variation analysis. The major and minor influencing factors for each region and each season are summarized in the following, along with the variation patterns of RH-$PM_{2.5}$ correlation, TEM-$PM_{2.5}$ correlation, WS-$PM_{2.5}$ correlation, and PS-$PM_{2.5}$ correlation. Fig. 3 shows the flow chart of the study design.

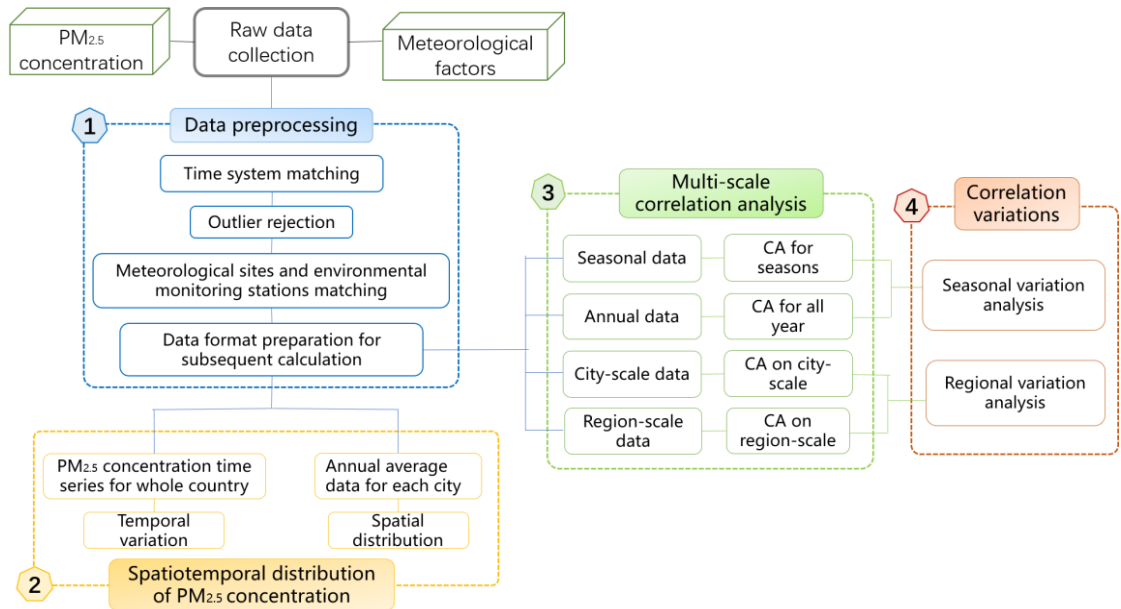

**Fig. 3.** Flow chart of the study design.

According to topography and climate, the division of the seven regions in China is as shown in Supplementary Fig. 1 and Supplementary Table 2. Considering the climate characteristics of China, the four seasons are: spring (March, April, May), summer (June, July, August), autumn (September, October, November), and winter (December, January, February).

## 3. Results and discussion

*3.1. Spatiotemporal variation of PM$_{2.5}$ concentration*

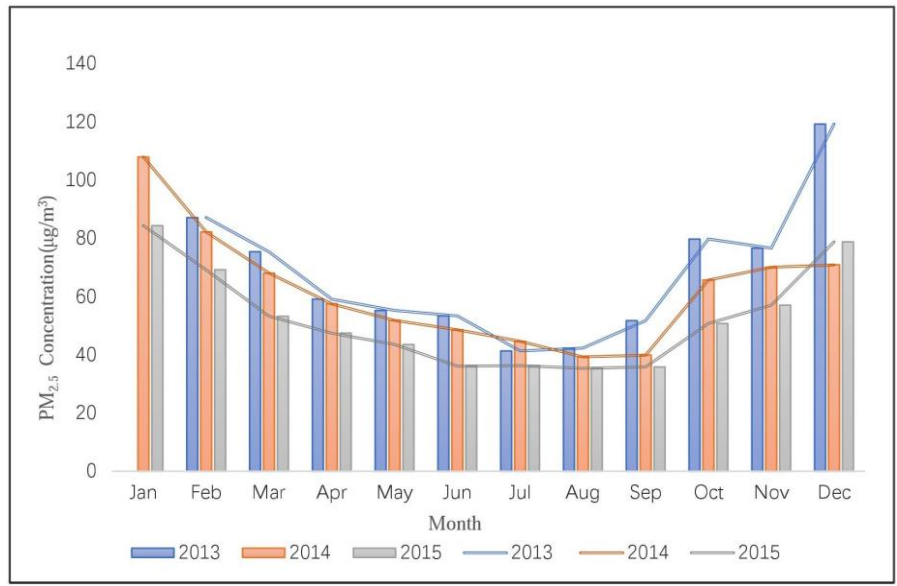

**Fig. 4.** Temporal variation of PM$_{2.5}$ concentration from 2013 to 2015.

Fig. 4 shows the monthly variation of China's average PM$_{2.5}$ concentration from 2013 to 2015. As shown in the figure, the PM$_{2.5}$ concentration in July and August is the lowest, and January and December have the highest concentrations. This means that the air pollution is much less serious in summer than in winter. This conclusion is similar to those of previous studies (Zhang et al., 2016; Yan et al., 2016). Another phenomenon worth noting in Fig. 4 is that the position of the blue line is at the top overall, while the grey line is at the bottom, which suggests that the PM$_{2.5}$ concentration of China has been decreasing from 2013 to 2015. This may indicate that the environmental protection policies and haze control measures in China have taken effect. Although the average air quality improved from 2013 to 2015, the number of cities that reached the Chinese Ambient Air Quality Standards (CAAQS) Grade II standard (35 $\mu$g/m$^3$) is still small, and the numbers are 5, 8, and 12, respectively, for 2013, 2014, and 2015.

We then analyzed the spatial distribution pattern of PM$_{2.5}$ concentration in 2015. Fig. 5(a) shows all the sites and the corresponding PM$_{2.5}$ concentration values we used to implement EBK, and Fig. 5(b) is the final interpolation result. It can be seen that the BTH region and Urumqi suffer from serious PM$_{2.5}$ pollution, but the air quality in Tibet, Yunnan province, Hainan Island, and the PRD area is better. The interpolation result is consistent with the results of previous studies (Ma et al., 2014; Geng et al., 2015; Li et al., 2017).

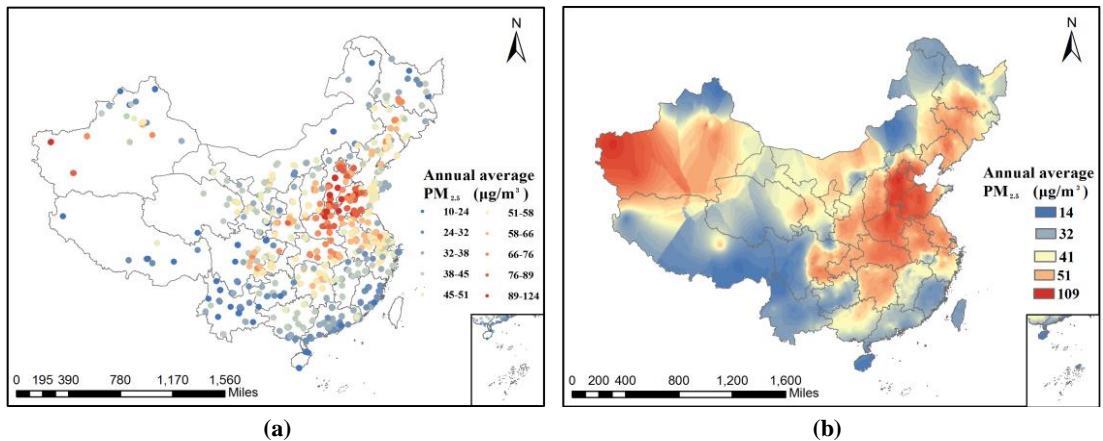

**Fig. 5.** (a) Annual PM$_{2.5}$ concentration at the monitoring stations in 2015. (b) The distribution of annual average PM$_{2.5}$ concentration in China interpolated by EBK.

*3.2. Regional variation of the correlation relationship*

*3.2.1 Correlation analysis at the city scale*

We calculated the Spearman's rank correlation coefficients between PM$_{2.5}$ concentration and RH, TEM, WS, and PS in the 68 cities. The result is provided in Supplementary Table 3. The correlation coefficients and p-values in the cities of South China and North China are listed in Table 1. In the table, r-RH, r-TEM, r-WS, and r-PS stand for the correlation coefficients between PM$_{2.5}$ concentration and RH, TEM, WS, and PS respectively, and the corresponding P-RH, P-TEM, P-WS, and P-PS stand for the p-values in the hypothesis testing of the correlation between PM$_{2.5}$ concentration and RH, TEM, WS, and PS. Most of the p-values in Supplementary Table 3 and Table 1 are less than 0.05, which indicates that $PM_{2.5}$ concentration is significantly correlated with meteorological factors, at a 95% confidence level. However, it is worth noting that the p-values in the last column are larger than the other columns, which indicates that the correlation between $PM_{2.5}$ concentration and pressure is not very significant in some cities.

Table 1. The correlation coefficients and p-values of cities in South China and North China.

| Region | City | Correlation coefficient (r) | | | | Significance p-value | | | |
|---|---|---|---|---|---|---|---|---|---|
| | | r-RH | r-TEM | r-WS | r-PS | P-RH | P-TEM | P-WS | P-PS |
| North China | Beijing | 0.484 | −0.072 | −0.376 | −0.004 | 0.0000 | 0.0009 | 0.0000 | 0.2176 |
| | Tianjin | 0.307 | −0.106 | −0.206 | −0.075 | 0.0000 | 0.0000 | 0.0000 | 0.0405 |
| | Shijiazhuang | 0.331 | −0.368 | −0.291 | −0.228 | 0.0000 | 0.0000 | 0.0000 | 0.0170 |
| | Tangshan | 0.294 | −0.149 | −0.202 | −0.204 | 0.0000 | 0.0000 | 0.0000 | 0.3530 |
| | Qinhuangdao | 0.161 | −0.202 | 0.022 | −0.190 | 0.0000 | 0.0000 | 0.4443 | 0.1493 |
| | Baoding | 0.272 | −0.380 | −0.212 | −0.067 | 0.0000 | 0.0000 | 0.0000 | 0.9763 |
| | Zhangjiakou | 0.166 | −0.300 | −0.030 | −0.030 | 0.0000 | 0.0000 | 0.0074 | 0.8171 |
| | Chengde | 0.238 | −0.137 | −0.221 | 0.034 | 0.0000 | 0.0002 | 0.0000 | 0.0696 |
| | Xingtai | 0.274 | −0.370 | −0.266 | 0.243 | 0.0000 | 0.0000 | 0.0000 | 0.0000 |
| | Taiyuan | 0.062 | −0.287 | −0.248 | 0.192 | 0.0104 | 0.0000 | 0.0000 | 0.0433 |
| | Huhehaote | 0.091 | −0.174 | −0.088 | 0.153 | 0.0481 | 0.0000 | 0.0133 | 0.0000 |
| South China | Guangzhou | −0.376 | −0.427 | −0.179 | 0.444 | 0.0000 | 0.0000 | 0.0037 | 0.0000 |
| | Shenzhen | −0.504 | −0.531 | −0.031 | −0.061 | 0.0000 | 0.0000 | 0.3108 | 0.0637 |
| | Zhuhai | −0.502 | −0.596 | −0.233 | 0.119 | 0.0000 | 0.0000 | 0.0000 | 0.0174 |
| | Foshan | −0.423 | −0.440 | −0.233 | 0.471 | 0.0000 | 0.0000 | 0.0000 | 0.0000 |
| | Zhongshan | −0.445 | −0.510 | −0.274 | 0.073 | 0.0000 | 0.0000 | 0.0000 | 0.0019 |
| | Dongguan | −0.375 | −0.450 | −0.273 | 0.153 | 0.0000 | 0.0000 | 0.0000 | 0.2288 |
| | Huizhou | −0.585 | −0.453 | −0.019 | −0.027 | 0.0000 | 0.0000 | 0.7030 | 0.5015 |
| | Zhaoqing | −0.322 | −0.453 | −0.356 | −0.188 | 0.0000 | 0.0000 | 0.0000 | 0.0005 |
| | Nanning | −0.324 | −0.534 | −0.435 | 0.592 | 0.0000 | 0.0000 | 0.0000 | 0.0000 |
| | Haikou | −0.157 | −0.590 | 0.220 | 0.606 | 0.0000 | 0.0000 | 0.0045 | 0.0000 |

In Table 1 and Supplementary Table 3, the regional variation can be clearly seen. For instance, $PM_{2.5}$ is positively correlated with RH in Beijing, with the correlation coefficient (r) value reaching 0.48, but in Shenzhen and Huizhou, $PM_{2.5}$ is negatively correlated with RH, and the correlation coefficients are −0.50 and −0.59, respectively. Furthermore, the correlation coefficients between $PM_{2.5}$ and PS are large in Central China, but relatively small in Northwest China. This result shows that spatial variations do exist in the relationships between $PM_{2.5}$ and meteorological factors.

To describe the regional variation in a more intuitive way, we display the correlation coefficients of each city on the map in Fig. 6, and represent the correlation coefficient (r) values with colored points. Green represents a negative correlation and red represents a positive correlation, with a deeper color indicating a more closely correlated relationship. Fig. 6(a) shows the correlation coefficients between $PM_{2.5}$ concentration and relative humidity in the 68 cities. The relationship is a positive correlation in North China and Urumqi, but is a negative correlation in other areas, especially South China. We infer that this variation may result from the type difference of the aerosol in North China and South China. North China is a region of heavy industry, and the common aerosol of this region is the City-Industry type (Barnaba et al., 2004), in which nitrate and sulfate account for a large proportion of the $PM_{2.5}$ (Cao et al., 2012; Wang et al., 2014). However, in South China, the aerosol type is Clean-Ocean type (Barnaba et al., 2014), and NaCl is one of the main components in the $PM_{2.5}$ of this region. According to related studies (Hu, et al., 2010; Ye et al., 2013), NaCl is more likely to absorb large amounts of moisture and fall to the ground under the high humidity situation of Southeast China, but the hygroscopicity of nitrate and sulfate is weak, and they just grow heavier rather than falling to the ground under the drier situation in North China. Furthermore, according to the study of Dawson et al. (2007), high humidity can benefit the formation of ammonium nitrate, further causing the positive correlation in North China.

Temperature has a strong negative correlation with $PM_{2.5}$ concentration in nearly all the cities of China, as shown in Fig. 6(b), and it can also reflect the seasonal variation in PM$_{2.5}$ concentration. PM$_{2.5}$ concentration is the lowest in summer and is the highest in winter, and temperature shows the opposite trend. The reason for the negative correlation may be that high temperature promotes the convection of air, and thus brings about the dilution and dispersion of air pollutants (Luo et al., 2017). And the low temperature can lead to increased emission rates from domestic heating and power production (Tran et al., 2011). In addition, our results indicate that the correlation between PM$_{2.5}$ concentration and TEM is weaker in North China than other areas.

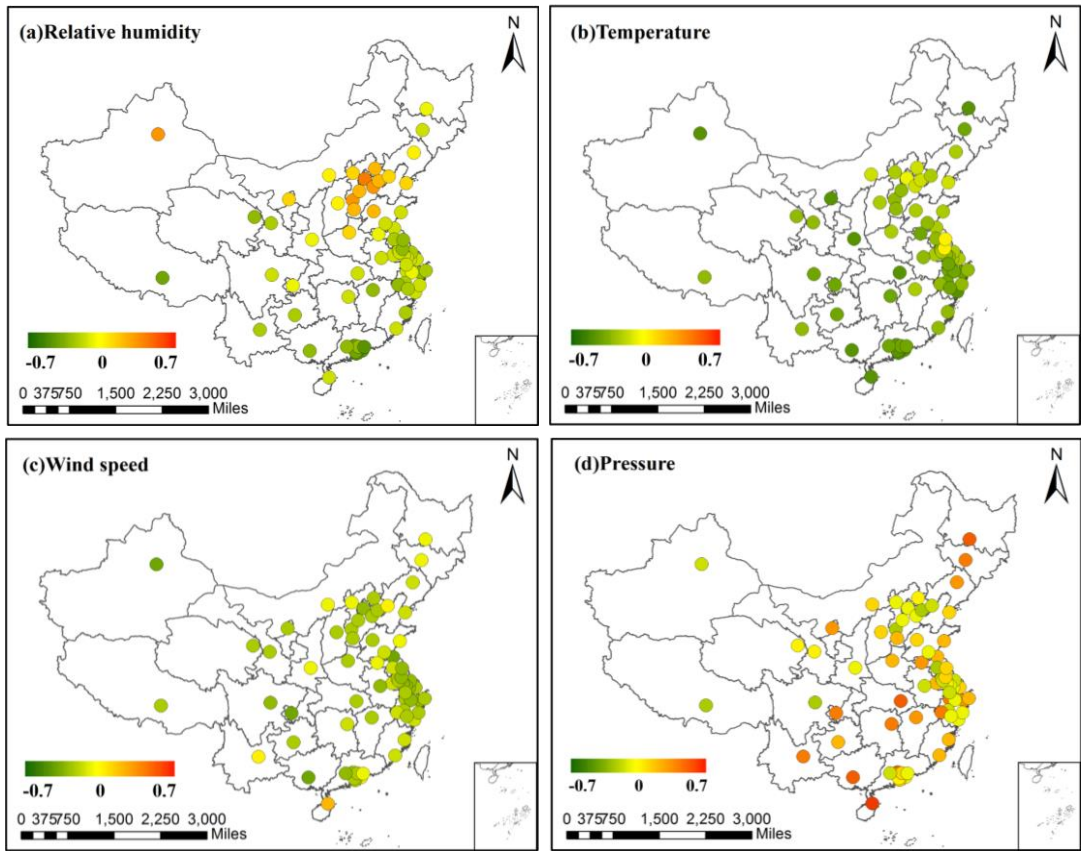

**Fig. 6.** The correlation coefficient (r) values in the 68 major cities. (a) Correlation between PM$_{2.5}$ concentration and RH. (b) Correlation between PM$_{2.5}$ concentration and TEM. (c) Correlation between PM$_{2.5}$ concentration and WS. (d) Correlation between PM$_{2.5}$ concentration and PS.

In Fig. 6(c), wind speed is negatively correlated with PM$_{2.5}$ concentration in most areas, which indicates that horizontal dispersion plays a significant role in China. Hainan province is an exception, with a positive PM$_{2.5}$-WS correlation, which is because the ambient air in Hainan Island is very clean (Li et al., 2017; Lu et al., 2017), and the wind from the surrounding areas can bring pollutants to the island, thus increasing the PM$_{2.5}$ concentration.

A positive correlation is found between PM$_{2.5}$ concentration and surface pressure in Northeast China, Central China, and Hainan province, as shown in Fig. 6(d). This is because high pressure is always accompanied with low atmospheric boundary layer height, and thus the vertical dispersion of air pollutants is hindered (Wu et al., 2017). The correlation in other areas is relatively weak. The reason for this regional difference is still unclear and needs further research.

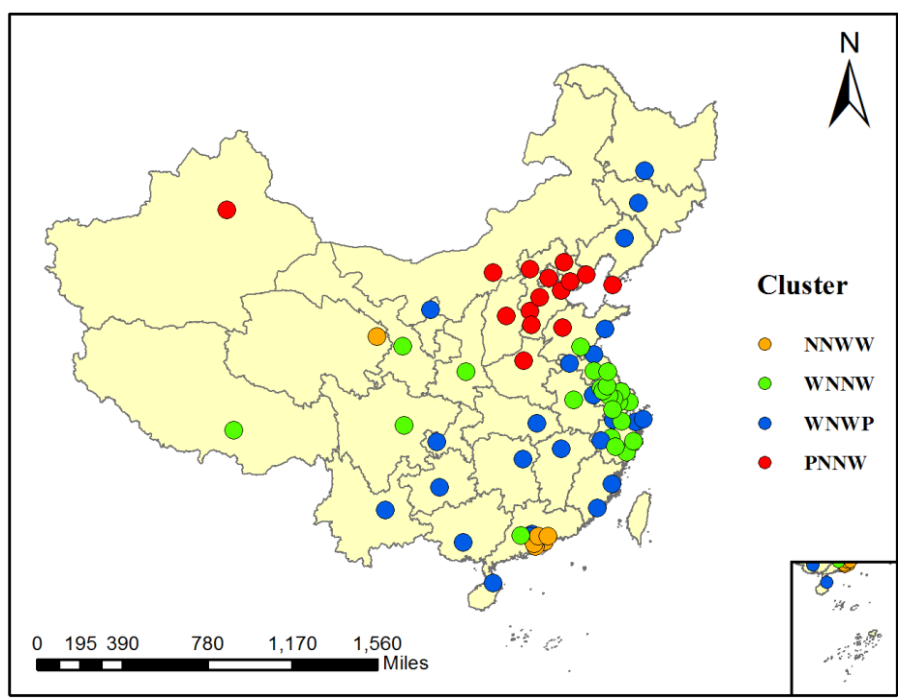

**Fig. 7.** Clustering result of the 68 cities. (The letter "N" in the cluster name represents "Negative correlation" with r<−0.2 or r=−0.2, "P" represents "Positive correlation" with r>0.2 or r=0.2, "W" represents "Weak correlation" with −0.2<r<0.2. And four letters of each cluster name stand for the correlations between PM$_{2.5}$ concentration and RH, TEM, WS, PS sequentially)

To understand the spatial variations more clearly, we also conducted a cluster analysis for the 68 cities. The ISODATA algorithm was used to implement the clustering. The clustering results are shown in Fig. 7, where the 68 cities are clustered into four groups. The PRD region is clustered as the first group, although with individual exceptions. In this area, $PM_{2.5}$ concentration is negatively correlated with RH and TEM, but weakly correlated with WS and PS. This cluster type is displayed by the orange points in Fig. 7. The YRD region and some cities in West China are clustered as the second type, symbolized by the green points, where $PM_{2.5}$ has a negative relationship with TEM and WS and a weak correlation with the other meteorological factors. The blue points represent the third type and mainly include cities in Northeast China and Central China, where $PM_{2.5}$ concentration is negatively correlated with TEM and positively correlated with PS. Lastly, North China and Urumqi are grouped as the fourth type, as shown by the red points, where $PM_{2.5}$ concentration in the cities of these regions is positively correlated with RH, negatively correlated with TEM and WS, and weakly correlated with PS.

The clustering result shows that there are both geographic dependencies and geographic differences in the relationships between $PM_{2.5}$ concentration and meteorological factors.

*3.2.2 Correlation analysis at the regional scale*

To obtain an exact evaluation of the regional difference of the correlations between $PM_{2.5}$ concentration and meteorological factors, we conducted a correlation analysis at a regional scale. As has been mentioned before, we divided mainland China into seven regions. Fig. 8 displays the regional correlation analysis results. In the figure, there is a histogram for each area, and the different colors of the bars represent the correlation coefficient (r) values of the different meteorological factors. An upward bar means a positive correlation and downward means negative. North China is the only area where $PM_{2.5}$ concentration is positively correlated with RH but, overall, $PM_{2.5}$ is only weakly correlated with meteorological factors in North China compared with the other areas. The most important influencing factors for PM$_{2.5}$ in South China are temperature and relative humidity. Surface pressure plays an important role in Central China and North China, in addition to temperature. As for East China, Southwest China, and Northwest China, temperature is the only factor that has correlation coefficients larger than 0.3 with PM$_{2.5}$ concentration.

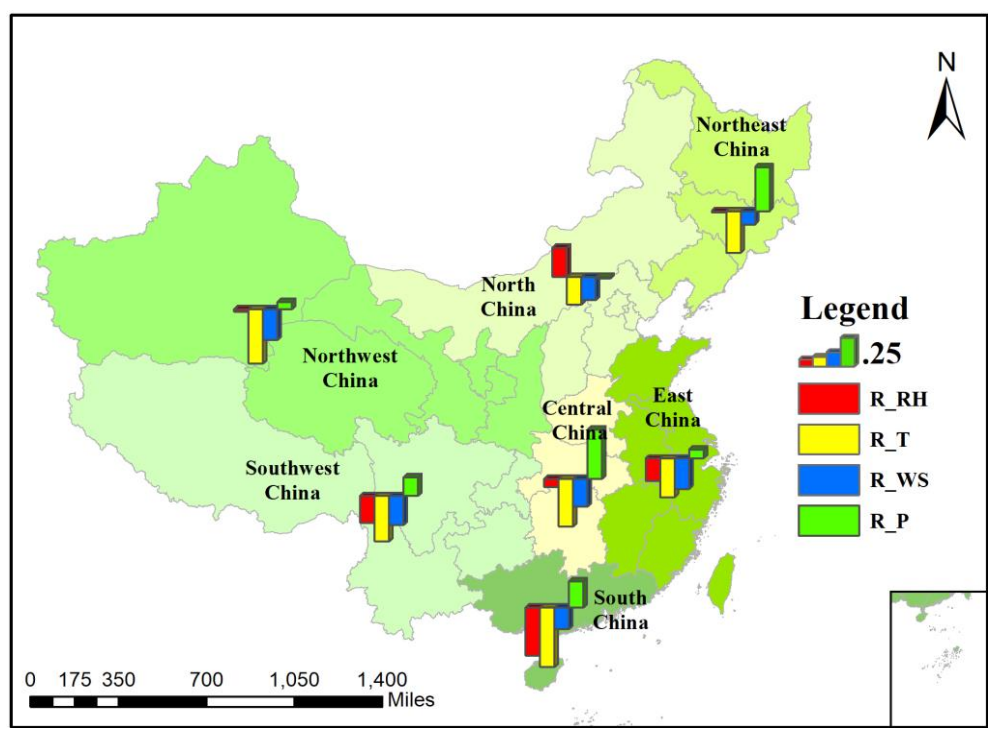

**Fig. 8.** Histograms of the correlation coefficient (r) values in the seven regions.

We summarize the major ($|r|>0.3$) and minor ($0.3>|r|>0.2$) influencing factors of the seven regions in Table 2. The seven regions are classified into four types according to their major and minor influencing factors, and the type numbers are marked behind the region names in the parentheses. We can see that this classification result is similar to the ISODATA clustering result in Fig. 7. As we can see, TEM is a major influencing factor for almost every region, but PS and RH mainly affect the PM$_{2.5}$ concentration in certain regions only.

Table 2. The major and minor influencing factors of the seven regions.

| Region (type number) | Major factors ($|r|>0.3$) | Minor factors ($0.3>|r|>0.2$) |
|---|---|---|
| North China (1) | \ | RH, TEM |
| South China (2) | TEM, RH | PS |
| Central China (3) | PS, TEM | WS |
| Northeast China (3) | PS, TEM | \ |
| Southwest China (4) | TEM | WS |
| Northwest China (4) | TEM | WS |
| East China (4) | TEM | WS |

*3.3. Seasonal variation of the correlation relationships*

The relationships between $PM_{2.5}$ concentration and meteorological factors not only vary with the geographic locations, but also with the seasons. Thus, we separated our data into four parts according to the season and conducted a correlation analysis in each of the four seasons. Fig. 9 shows the seasonal variations of the correlation coefficient (r) values of $PM_{2.5}$ and the four meteorological factors.

As shown in Fig. 9, the relationship between $PM_{2.5}$ and RH varies only slightly with the seasons, and the positive correlation in spring and winter is only slightly stronger than in summer and autumn. The number of cities with a positive correlation is also larger in spring and winter. We infer that this is the result of the increase of coal consumption in the cold season. Increased coal combustion can result in increased emission of $SO_2$ and other air pollutants, which can promote the formation of sulfate, to a certain extent (Zheng et al., 2005; Dan et al., 2004). As mentioned before, the sulfate in $PM_{2.5}$ is one of the most important components that results in the positive relationship between $PM_{2.5}$ and RH. Therefore, the increase in the sulfate content makes the positive correlation stronger.

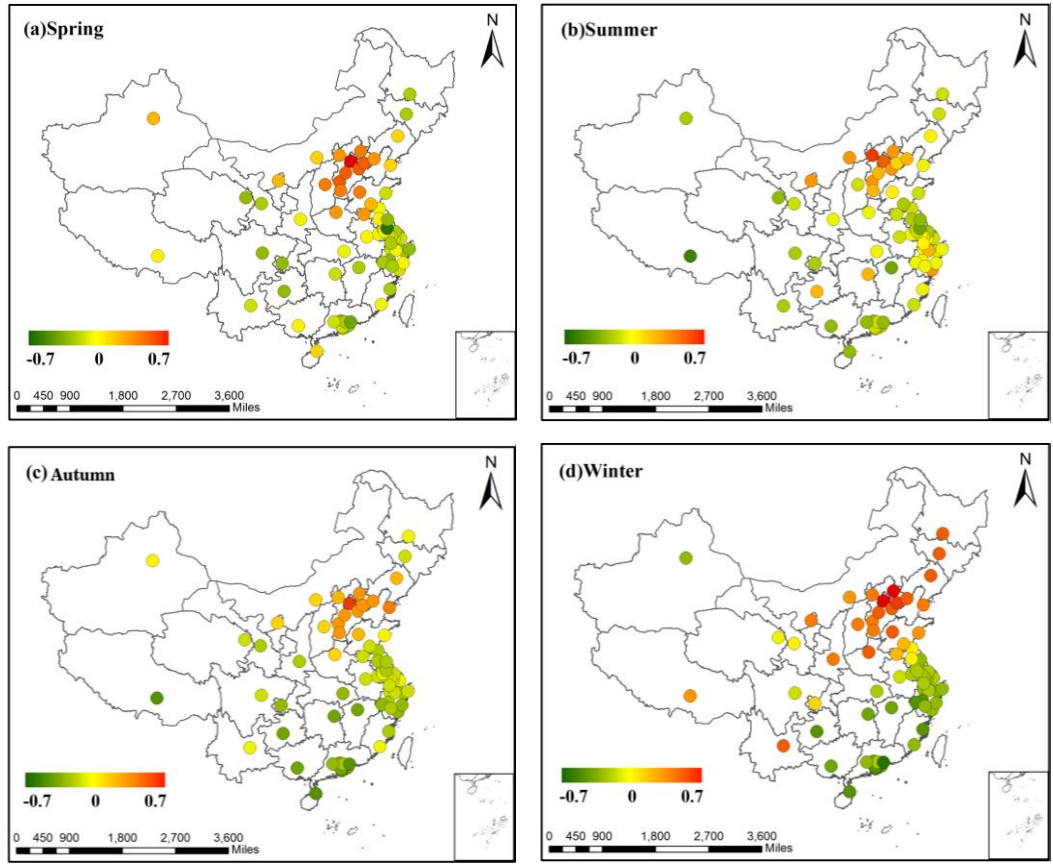

**Fig. 9.** The distribution of the correlation coefficient (r) values between $PM_{2.5}$ and RH in the four seasons.

In Fig. 10, the correlation between $PM_{2.5}$ and TEM varies greatly with the seasons. Overall, $PM_{2.5}$ is positively correlated with TEM in winter and negatively correlated in autumn. In spring, the $PM_{2.5}$ concentration is negatively correlated with TEM in South China and Urumqi and is weakly correlated with TEM in the other cities. We can also see that there are two points in bright red color in East China in Fig. 10(a), but these results are not accurate because the sample quantity is too small. Due to missing data in these two cities, the $PM_{2.5}$ concentration data and corresponding meteorological parameters used in the correlation analysis in spring are less than 10 pairs. In summer, the situation is the opposite to that in spring. Weak positive correlation is found in South China, North China, Northeast China, Southwest China, and Northwest China, and the correlation between $PM_{2.5}$ and TEM is very weak in Central China and East China.

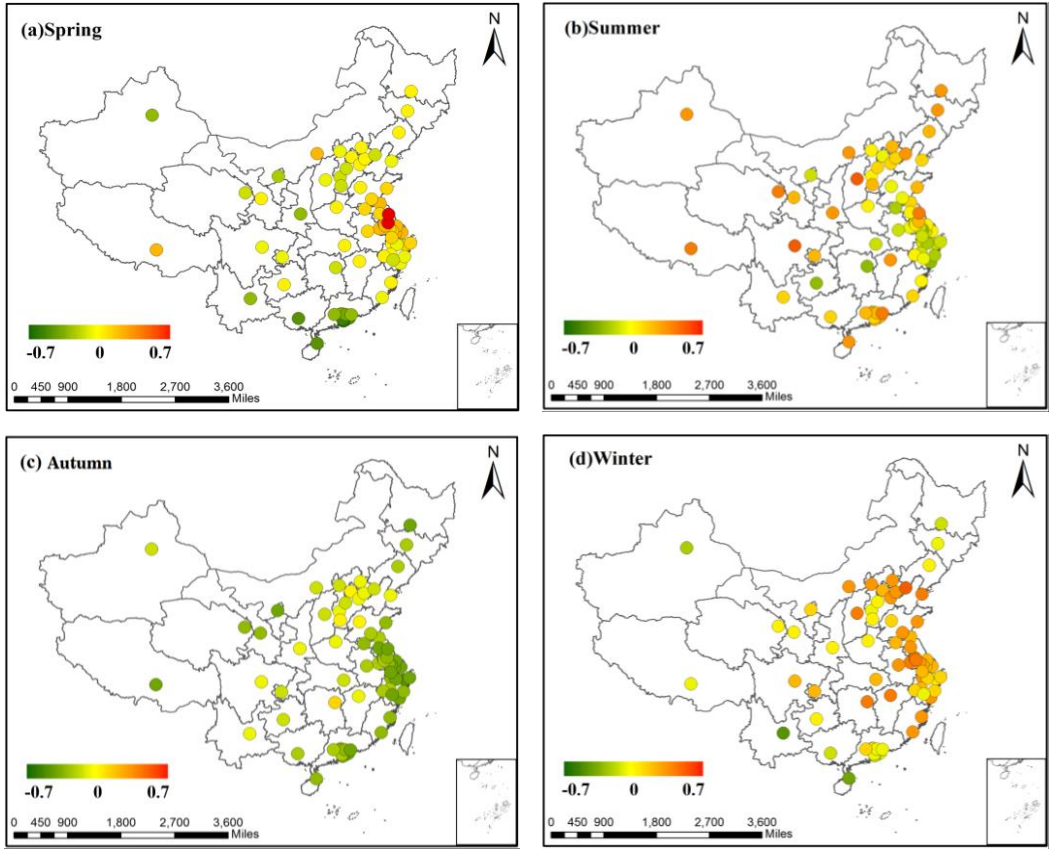

**Fig. 10.** The distribution of the correlation coefficient (r) values between $PM_{2.5}$ and TEM in the four seasons.

In Fig. 11, the relationship between WS and $PM_{2.5}$ is stable, and negative correlation is detected in each season, but the correlation turns weak in summer in some cities. We infer that this is because the atmospheric boundary layer is high in summer and the air convection in the vertical direction is strong due to the elevated temperature (Tai et al., 2010). Particles are mainly diffused and diluted vertically, so the correlation with the horizonal diffusion impact factor—wind speed—is reduced. The correlation in Hainan Island also turns to negative in summer, while the correlation coefficient (r) value is always positive in the other seasons. We speculate that this is because wind from sea to land prevails in summer (Wang et al., 2001), which promotes the diffusion and dilution of pollutants.

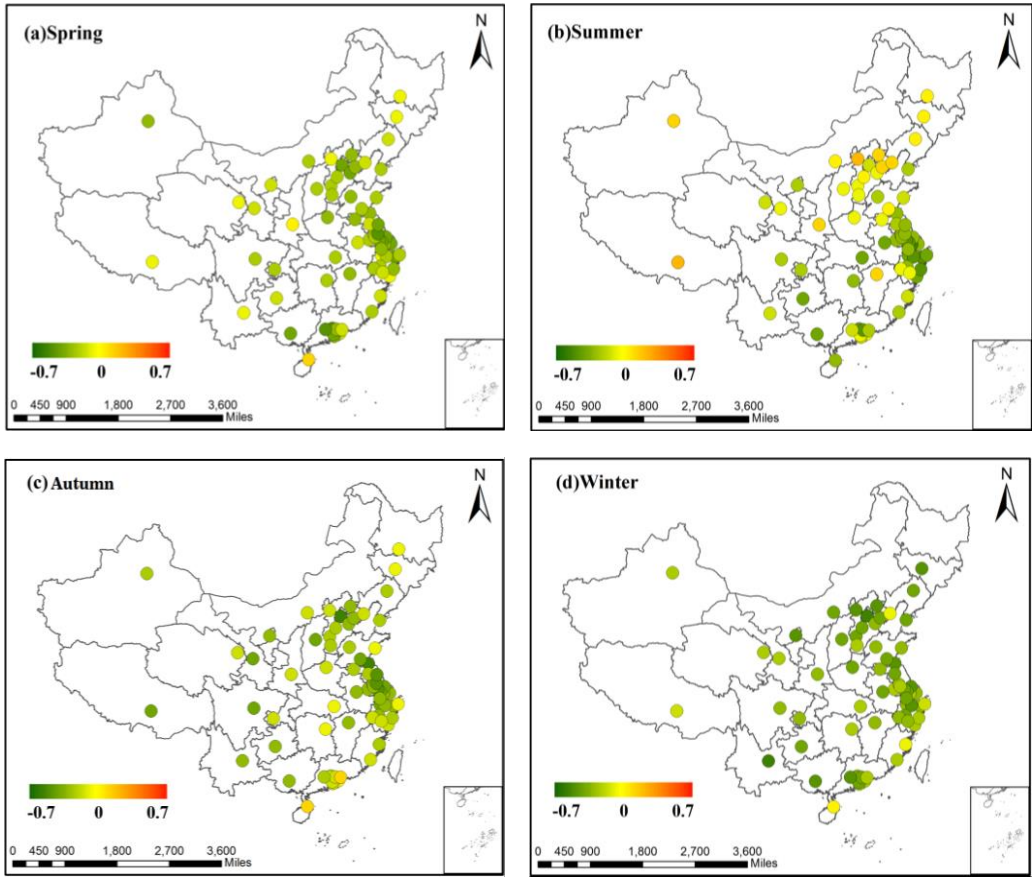

**Fig. 11.** The distribution of the correlation coefficient (r) values between $PM_{2.5}$ and WS in the four seasons.

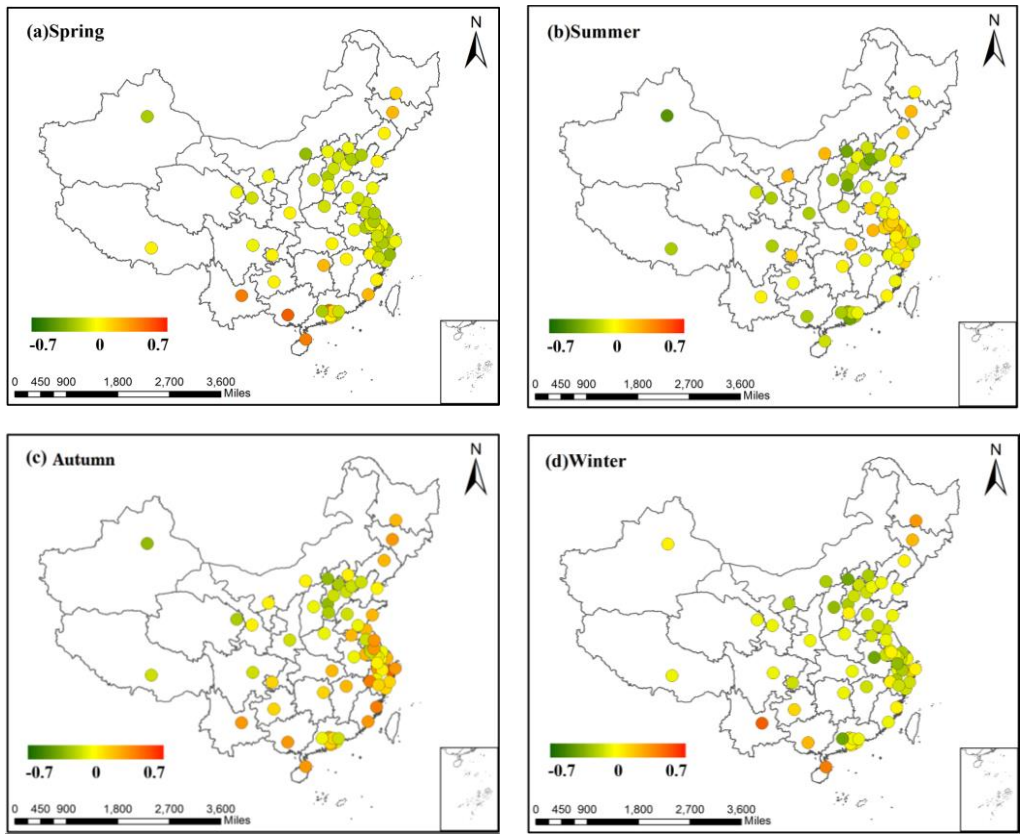

**Fig. 12.** The distribution of the correlation coefficient (r) values between $PM_{2.5}$ and PS in the four seasons.

Lastly, the result for pressure is displayed in Fig. 12. The seasonal variation of the relationship between $PM_{2.5}$ concentration and pressure is extreme, but the variation pattern is not obvious. Overall, the positive correlation in autumn is stronger than the other seasons, and the correlation coefficient distributions are similar in spring and winter.

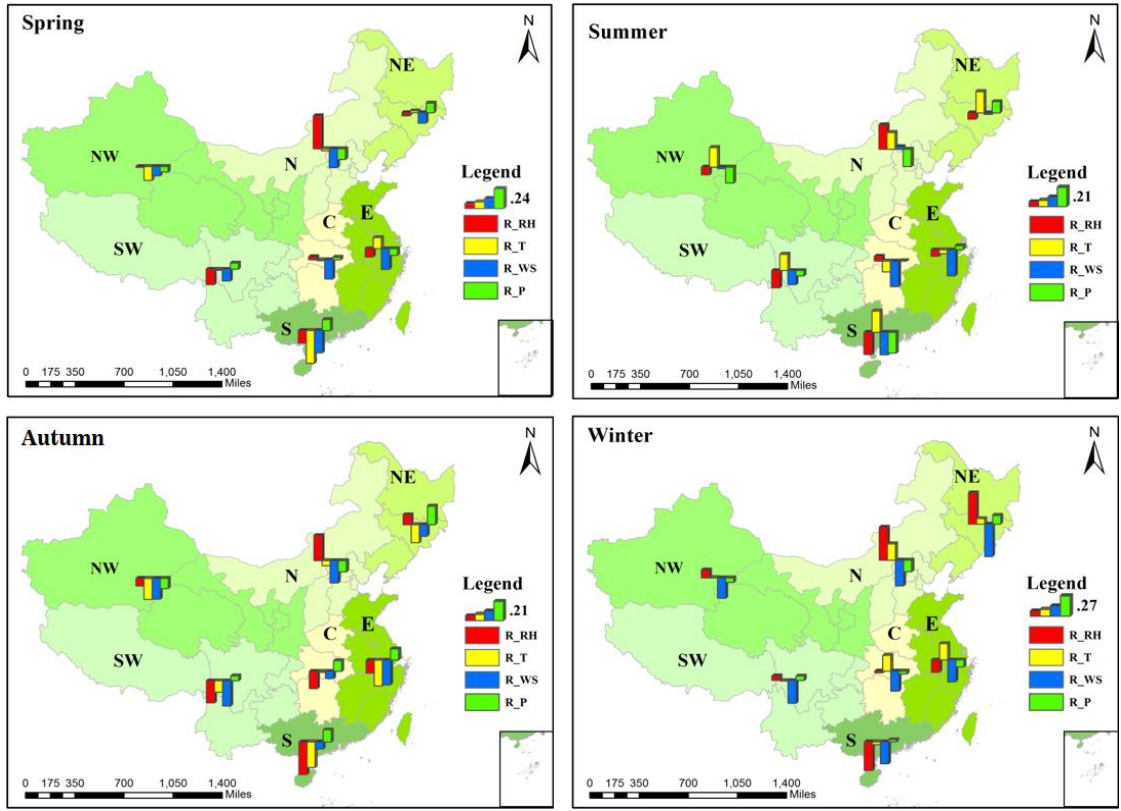

**Fig. 13.** Seasonal distribution of the correlation coefficient (r) value histograms in the seven regions.

In addition to the analysis at the city scale, we also conducted a regional correlation analysis in the four seasons. Fig. 13 shows the regional analysis result, and the correlation coefficient (r) values are displayed in Table 3 in detail. The color of each cell represents the sign and magnitude of the correlation coefficient (r) value. $PM_{2.5}$ concentration has the strongest correlation with meteorological factors in winter, and the correlation is the weakest in summer. This result is similar to the conclusion of a previous study (Chen et al., 2017). The major influencing factors of the four seasons are WS in spring, TEM in summer, TEM and WS in autumn, and WS in winter. The strong correlation between WS and $PM_{2.5}$ in winter may be due to the low atmospheric boundary layer height and the dependence on horizonal diffusion.

Table 3. The major and minor influencing factors of the seven regions (the color of each cell stands for the sign and magnitude of the correlation coefficient (r) value, where red stands for a positive correlation and blue stands for negative).

| | Spring | | | | | Summer | | | |
|---|---|---|---|---|---|---|---|---|---|
| **Region** | **RH** | **TEM** | **WS** | **PS** | **Region** | **RH** | **TEM** | **WS** | **PS** |
| Central China | 0.04 | −0.01 | −0.28 | 0.04 | Central China | 0.06 | −0.13 | −0.32 | 0.00 |
| North China | 0.48 | −0.03 | −0.28 | −0.16 | North China | 0.30 | 0.21 | 0.03 | −0.22 |
| South China | −0.18 | −0.48 | −0.33 | 0.16 | South China | −0.27 | 0.26 | −0.28 | −0.26 |
| Southwest China | −0.22 | −0.02 | −0.17 | 0.09 | Southwest China | −0.21 | 0.20 | −0.18 | −0.07 |
| Northwest China | −0.01 | −0.20 | −0.14 | −0.08 | Northwest China | −0.10 | 0.24 | −0.02 | −0.20 |
| Northeast China | −0.05 | 0.03 | −0.15 | 0.14 | Northeast China | −0.08 | 0.26 | −0.02 | 0.14 |
| East China | −0.12 | 0.16 | −0.30 | −0.10 | East China | −0.08 | −0.05 | −0.32 | 0.05 |
| **Average** | **−0.01** | **−0.08** | **−0.24** | **0.01** | **Average** | **−0.05** | **0.14** | **−0.16** | **−0.08** |
| | Autumn | | | | | Winter | | | |
| **Region** | **RH** | **TEM** | **WS** | **PS** | **Region** | **RH** | **TEM** | **WS** | **PS** |
| Central China | −0.23 | −0.01 | −0.10 | 0.14 | Central China | −0.03 | 0.26 | −0.33 | −0.05 |
| North China | 0.34 | −0.07 | −0.30 | −0.15 | North China | 0.54 | 0.27 | −0.43 | −0.19 |
| South China | −0.44 | −0.35 | −0.10 | 0.16 | South China | −0.48 | −0.05 | −0.37 | 0.03 |
| Southwest China | −0.30 | −0.15 | −0.34 | 0.07 | Southwest China | 0.08 | −0.01 | −0.38 | 0.07 |
| Northwest China | −0.10 | −0.29 | −0.28 | −0.14 | Northwest China | 0.13 | 0.00 | −0.34 | −0.07 |
| Northeast China | 0.13 | −0.25 | −0.16 | 0.25 | Northeast China | 0.52 | 0.09 | −0.54 | 0.14 |
| East China | −0.17 | −0.35 | −0.33 | 0.15 | East China | −0.20 | 0.27 | −0.36 | −0.12 |
| **Average** | **−0.11** | **−0.21** | **−0.23** | **0.07** | **Average** | **0.08** | **0.12** | **−0.39** | **−0.03** |

*3.4. Discussion*

Many scholars have worked on research into $PM_{2.5}$ concentration retrieval (Chu et al., 2016; Ma et al.,2016). Among all the existing methods, models that consider the spatiotemporal variations usually obtain higher retrieval accuracies. Taking the geographically weighted regression (GWR) model (Hu et al., 2013, Ma et al., 2014) as an example, it considers the spatiotemporal variations through changing the coefficients of the auxiliary variables in different areas and seasons, rather than using one model for all circumstances. For this type of retrieval model, the results of our work could provide some suggestions for the selection of auxiliary variables in different regions and seasons. For example, the correlation coefficient (r) value between $PM_{2.5}$ and PS is nearly zero in summer, so we can suggest that this factor could be excluded in the Central China summer model. The change of auxiliary variable type in the retrieval model could improve the $PM_{2.5}$ prediction accuracy.

Another phenomenon worth noting is that $PM_{2.5}$ concentration has a strong correlation with TEM in the analysis at a yearly scale (i.e., the analysis including the data of every season), but in the seasonal analysis, the correlation coefficient (r) values change to small values. We call this phenomenon "scale variation", which indicates that the correlation between $PM_{2.5}$ and meteorological factors varies with the temporal scale. For instance, in most of the long-time series (more than 10 years) analyses, the correlation between $PM_{2.5}$ and TEM is positive (Tai et al., 2010; Lu et al., 2017), but in most of the 1–2 years analyses (Li et al., 2015), the relationship turns to negative, and when it comes to the seasonal analysis, the result changes again (Huang et al., 2015; Zhang et al., 2015). We believe that this is because the long-time series analyses mainly reflect the year-to-year variation, and the yearly analyses mainly reflect the variation within the seasons. Therefore, the results of our analyses in the four seasons and over a whole year can be seen as evidence of the scale variation, which needs to be noted in future studies.

## 4. Conclusions

Spatial and temporal heterogeneity is one of the most important properties of $PM_{2.5}$ pollution, but many studies have ignored this aspect. In this study, through analyses of 68 cities and seven regions in four seasons during a time period of two years, we have proved the existence of regional and seasonal variations of the correlations between $PM_{2.5}$ concentration and meteorological factors.

(a) Spatially, the correlations between $PM_{2.5}$ concentration and meteorological factors would vary with regions. The evidence is that RH is positively correlated with $PM_{2.5}$ concentration in North China, but negatively correlated with $PM_{2.5}$ in South China and other areas. The positive correlation between TEM and $PM_{2.5}$ is weaker in North China than other areas. WS is negatively correlated with $PM_{2.5}$ in almost every region, expect for Hainan Island. PS has a strong positive correlation with $PM_{2.5}$ in Northeast China and Central China, but the correlation is weak in other places. The type of aerosol, the terrain, and the local climate can all be inducers of the regional variations.

(b) Seasonally, there exists seasonal variations of the correlation between $PM_{2.5}$ concentration and meteorological factors. The positive correlation between RH and $PM_{2.5}$ is stronger in winter and spring; TEM is positively correlated with $PM_{2.5}$ in winter and is negatively correlated with $PM_{2.5}$ in autumn; WS has the strongest correlation with $PM_{2.5}$ in winter; and the correlation between $PM_{2.5}$ and PS is the strongest in autumn. All the anthropogenic and natural differences, such as the use of heating systems in the North China winter, and weather variations in the four seasons, may bring about seasonal variations.

In this study, we investigated the relationships between $PM_{2.5}$ concentration and meteorological factors in terms of regional and seasonal variations, and we generated maps showing the varying patterns of the relationships. In this way, we were able to obtain a more comprehensive and precise understanding of how $PM_{2.5}$ concentration is correlated with meteorological factors. This knowledge could provide a solid foundation for more accurate $PM_{2.5}$ concentration retrieval, and for making more effective environmental protection policies for different regions. Although lots of work has been done in this area of research, there is still much room for improvement. Firstly, a global analysis with a longer time series and more factors, such as precipitation, sunlight duration, and terrain, would enable us to explain the variation more thoroughly. Secondly, there are still some phenomena that we cannot fully explain, and for which further research is needed, such as the regional variation of the $PM_{2.5}$-PS correlation. The exploration of the cause of this regional and seasonal variation would help us to better understand the air pollution problem in China.

Despite these limitations, our work could still provide a better chance for a more accurate forecast of $PM_{2.5}$ concentration. Through considering the spatial and seasonal variations of the relationships between $PM_{2.5}$ concentration and meteorological factors, the retrieval model of $PM_{2.5}$ concentration could be more precise at a finer scale, and the concomitant improvement of haze forecast accuracy would be of great benefit.

**Acknowledgments**

This work was supported by the National Key R&D Program of China (no. 2016YFC0200900). We would also like to thank the data providers of the China National Environmental Monitoring Center (CNEMC) and the China Meteorological Data Service Center (CMDC).

- **Supplementary materials**

Supplementary Table 1. The result of the parameter sensitivity test in Beijing.

| Buffer radius | r-RH | r-TEM | r-WS | r-PS | City number |
|---|---|---|---|---|---|
| 0.1° | 0.49 | −0.13 | −0.19 | 0.05 | 24 |
| 0.2° | 0.49 | −0.13 | −0.20 | 0.05 | 44 |
| 0.3° | 0.45 | −0.11 | −0.24 | 0.03 | 68 |
| 0.4° | 0.46 | −0.11 | −0.24 | 0.04 | 72 |
| 0.5° | 0.46 | −0.13 | −0.33 | 0.06 | 74 |
| 0.6° | 0.44 | −0.13 | −0.32 | 0.06 | 74 |

City number is the number of cities in which $PM_{2.5}$ concentration data and meteorological parameters were successfully matched. r-RH, r-TEM, r-WS, and r-PS represent the correlation coefficients between $PM_{2.5}$ concentration and RH, TEM, WS, and PS.

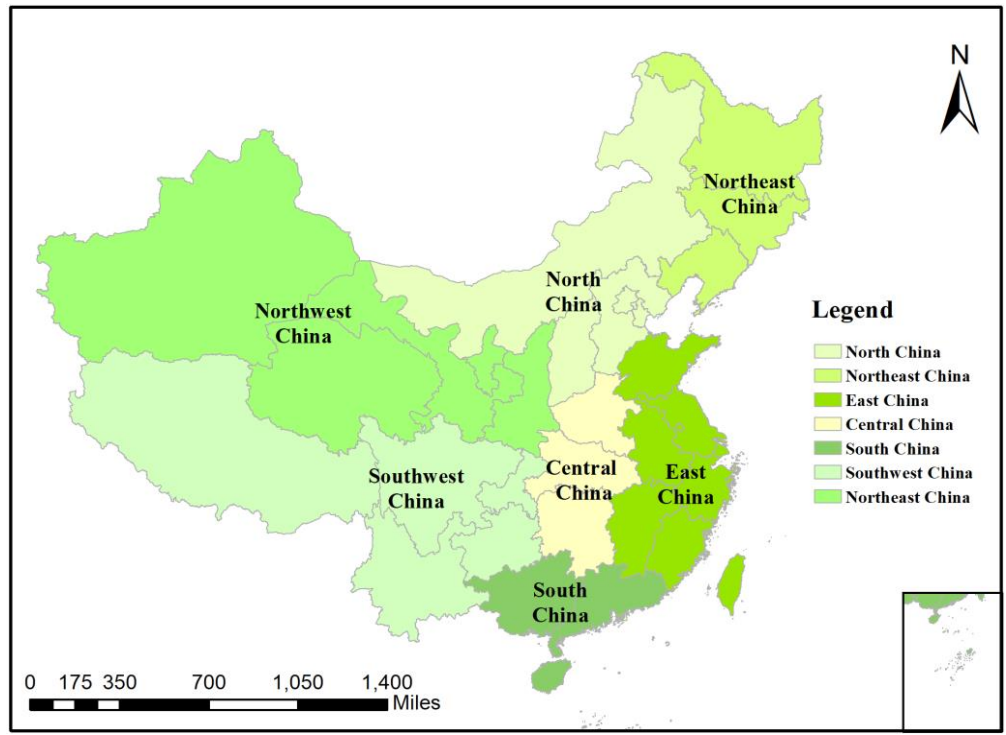

**Supplementary Fig. 1.** The seven regions in China.

Supplementary Table 2. The corresponding relationships between cities, provinces, and regions.

| Region | Province | City | Region | Province | City |
|---|---|---|---|---|---|
| **North China** | Beijing | Beijing | **East China** | Jiangsu | Nanjing |
| | Tianjin | Tianjin | | | Shanghai |
| | Hebei | Shijiazhuang | | | Suzhou |
| | | Tangshan | | | Nantong |
| | | Qinhuangdao | | | Lianyungang |
| | | Baoding | | | Xuzhou |
| | | Zhangjiakou | | | Yangzhou |
| | | Chengde | | | Wuxi |
| | | Xingtai | | | Changzhou |
| | Shanxi | Taiyuan | | | Zhenjiang |
| | Nei Mongolia | Huhehaote | | | Taizhou |
| **South China** | Guangdong | Guangzhou | | | Huai'an |
| | | Shenzhen | | | Yancheng |
| | | Zhuhai | | | Suqian |
| | | Foshan | | Zhejiang | Hangzhou |
| | | Zhongshan | | | Ningbo |
| | | Dongguan | | | Wenzhou |
| | | Huizhou | | | Shaoxing |
| | | Zhaoqing | | | Huzhou |
| | Guangxi | Nanning | | | Taizhou |
| | Hainan | Haikou | | | Zhoushan |
| **Central China** | Hubei | Wuhan | | | Jinhua |
| | Henan | Zhengzhou | | | Quzhou |
| | Hunan | Changsha | | | Lishui |
| **Northwest China** | Shaanxi | Xi'an | | Anhui | Hefei |
| | Gansu | Lanzhou | | Fujian | Fuzhou |
| | Qinghai | Xining | | | Xiamen |
| | Ningxia | Yinchuan | | Jiangxi | Nanchang |
| | Xinjiang | Urumqi | | Shandong | Jinan |
| **Northeast China** | Chongqing | Chongqing | | | Qingdao |
| | Sichuan | Chengdu | **Northeast China** | Liaoning | Shengyang |
| | Guizhou | Guiyang | | | Dalian |
| | Yunnan | Kunming | | Jilin | Changchun |
| | Xizang | Lhasa | | Heilongjiang | Ha'erbin |

Supplementary Table 3. The correlation coefficient (r) values and p-values between PM$_{2.5}$ concentration and the four meteorological factors in the 68 cities.

| Region | City | r-RH | P-RH | r-TEM | P-TEM | r-WS | P-WS | r-PS | P-PS |
|---|---|---|---|---|---|---|---|---|---|
| Northeast China | Shengyang | 0.061 | 0.007 | −0.299 | 0.000 | −0.145 | 0.000 | 0.353 | 0.000 |
| | Dalian | 0.121 | 0.002 | −0.179 | 0.000 | −0.173 | 0.000 | 0.134 | 0.000 |
| | Changchun | −0.148 | 0.728 | −0.401 | 0.000 | −0.043 | 0.001 | 0.481 | 0.000 |
| | Ha'erbin | −0.081 | 0.083 | −0.514 | 0.000 | −0.090 | 0.000 | 0.502 | 0.000 |
| North China | Beijing | 0.484 | 0.000 | −0.072 | 0.001 | −0.376 | 0.000 | −0.004 | 0.218 |
| | Tianjin | 0.307 | 0.000 | −0.106 | 0.000 | −0.206 | 0.000 | −0.075 | 0.041 |
| | Shijiazhuang | 0.331 | 0.000 | −0.368 | 0.000 | −0.291 | 0.000 | −0.228 | 0.017 |
| | Tangshan | 0.294 | 0.000 | −0.149 | 0.000 | −0.202 | 0.000 | −0.204 | 0.353 |
| | Qinhuangdao | 0.161 | 0.000 | −0.202 | 0.000 | 0.022 | 0.444 | −0.190 | 0.149 |
| | Baoding | 0.272 | 0.000 | −0.380 | 0.000 | −0.212 | 0.000 | −0.067 | 0.976 |
| | Zhangjiakou | 0.166 | 0.000 | −0.300 | 0.000 | −0.030 | 0.007 | −0.030 | 0.817 |
| | Chengde | 0.238 | 0.000 | −0.137 | 0.000 | −0.221 | 0.000 | 0.034 | 0.070 |
| | Xingtai | 0.274 | 0.000 | −0.370 | 0.000 | −0.266 | 0.000 | 0.243 | 0.000 |
| | Taiyuan | 0.062 | 0.010 | −0.287 | 0.000 | −0.248 | 0.000 | 0.192 | 0.043 |
| | Huhehaote | 0.091 | 0.048 | −0.174 | 0.000 | −0.088 | 0.013 | 0.153 | 0.000 |
| South China | Guangzhou | −0.376 | 0.000 | −0.427 | 0.000 | −0.179 | 0.004 | 0.444 | 0.000 |
| | Shenzhen | −0.504 | 0.000 | −0.531 | 0.000 | −0.031 | 0.311 | −0.061 | 0.064 |
| | Zhuhai | −0.502 | 0.000 | −0.596 | 0.000 | −0.233 | 0.000 | 0.119 | 0.017 |
| | Foshan | −0.423 | 0.000 | −0.440 | 0.000 | −0.233 | 0.000 | 0.471 | 0.000 |
| | Zhongshan | −0.445 | 0.000 | −0.510 | 0.000 | −0.274 | 0.000 | 0.073 | 0.002 |
| | Dongguan | −0.375 | 0.000 | −0.450 | 0.000 | −0.273 | 0.000 | 0.153 | 0.229 |
| | Huizhou | −0.585 | 0.000 | −0.453 | 0.000 | −0.019 | 0.703 | −0.027 | 0.501 |
| | Zhaoqing | −0.322 | 0.000 | −0.453 | 0.000 | −0.356 | 0.000 | −0.188 | 0.001 |
| | Nanning | −0.324 | 0.000 | −0.534 | 0.000 | −0.435 | 0.000 | 0.592 | 0.000 |
| | Haikou | −0.157 | 0.000 | −0.590 | 0.000 | 0.220 | 0.005 | 0.606 | 0.000 |
| Central China | Wuhan | −0.197 | 0.001 | −0.510 | 0.000 | −0.249 | 0.000 | 0.518 | 0.000 |
| | Zhengzhou | 0.143 | 0.000 | −0.282 | 0.000 | −0.255 | 0.000 | 0.230 | 0.000 |
| | Changsha | −0.152 | 0.002 | −0.408 | 0.000 | −0.188 | 0.018 | 0.483 | 0.000 |
| Northwest China | Xi'an | −0.053 | 0.157 | −0.509 | 0.000 | −0.085 | 0.000 | −0.018 | 0.208 |
| | Lanzhou | −0.212 | 0.000 | −0.302 | 0.000 | −0.255 | 0.009 | 0.087 | 0.031 |
| | Xining | −0.359 | 0.000 | −0.399 | 0.000 | −0.255 | 0.002 | 0.049 | 0.268 |
| | Yinchuan | 0.134 | 0.000 | −0.556 | 0.000 | −0.294 | 0.000 | 0.363 | 0.000 |
| | Urumqi | 0.374 | 0.000 | −0.503 | 0.000 | −0.410 | 0.000 | −0.189 | 0.010 |

| Region | City | r-RH | P-RH | r-T | P-T | r-WS | P-WS | r-P | P-P |
|---|---|---|---|---|---|---|---|---|---|
| Southwest China | Chongqing | −0.008 | 0.018 | −0.462 | 0.000 | −0.446 | 0.000 | 0.471 | 0.000 |
| | Chengdu | −0.170 | 0.001 | −0.416 | 0.000 | −0.339 | 0.000 | −0.224 | 0.000 |
| | Guiyang | −0.237 | 0.000 | −0.408 | 0.000 | −0.248 | 0.000 | 0.298 | 0.000 |
| | Kunming | −0.277 | 0.000 | −0.309 | 0.000 | 0.070 | 0.382 | 0.436 | 0.000 |
| | Lhasa | −0.429 | 0.000 | −0.300 | 0.000 | −0.242 | 0.000 | −0.201 | 0.000 |
| East China | Nanjing | −0.174 | 0.005 | −0.296 | 0.000 | −0.179 | 0.000 | 0.268 | 0.000 |
| | Shanghai | −0.174 | 0.000 | −0.252 | 0.000 | −0.247 | 0.000 | 0.156 | 0.000 |
| | Suzhou | −0.159 | 0.000 | −0.293 | 0.000 | −0.320 | 0.000 | −0.064 | 0.818 |
| | Nantong | −0.283 | 0.000 | −0.226 | 0.000 | −0.319 | 0.000 | −0.145 | 0.131 |
| | Lianyungang | −0.185 | 0.036 | −0.271 | 0.000 | −0.436 | 0.000 | 0.203 | 0.000 |
| | Xuzhou | −0.097 | 0.809 | −0.420 | 0.000 | −0.087 | 0.023 | 0.342 | 0.000 |
| | Yangzhou | −0.160 | 0.004 | −0.347 | 0.000 | −0.174 | 0.000 | −0.170 | 0.750 |
| | Wuxi | −0.103 | 0.051 | −0.357 | 0.000 | −0.346 | 0.000 | −0.063 | 0.876 |
| | Changzhou | −0.152 | 0.003 | −0.349 | 0.000 | −0.321 | 0.000 | −0.081 | 0.858 |
| | Zhenjiang | −0.174 | 0.000 | −0.242 | 0.000 | −0.264 | 0.000 | −0.159 | 0.708 |
| | Taizhou | −0.392 | 0.000 | 0.066 | 0.849 | −0.359 | 0.000 | 0.103 | 0.415 |
| | Huai'an | −0.218 | 0.028 | −0.363 | 0.000 | −0.122 | 0.001 | −0.206 | 0.046 |
| | Yancheng | −0.323 | 0.000 | 0.060 | 0.752 | −0.397 | 0.000 | 0.107 | 0.396 |
| | Suqian | −0.108 | 0.557 | −0.301 | 0.000 | −0.160 | 0.000 | −0.098 | 0.051 |
| | Hangzhou | −0.142 | 0.125 | −0.402 | 0.000 | −0.291 | 0.000 | 0.409 | 0.000 |
| | Ningbo | −0.312 | 0.000 | −0.447 | 0.000 | −0.376 | 0.000 | 0.388 | 0.000 |
| | Wenzhou | −0.157 | 0.000 | −0.532 | 0.000 | −0.177 | 0.000 | −0.091 | 0.004 |
| | Shaoxing | −0.021 | 0.674 | −0.459 | 0.000 | −0.326 | 0.000 | −0.089 | 0.165 |
| | Huzhou | −0.183 | 0.009 | −0.430 | 0.000 | −0.234 | 0.000 | −0.170 | 0.010 |
| | Taizhou | −0.151 | 0.000 | −0.374 | 0.000 | −0.244 | 0.000 | −0.091 | 0.084 |
| | Zhoushan | −0.282 | 0.000 | −0.381 | 0.000 | −0.296 | 0.000 | 0.289 | 0.000 |
| | Jinhua | −0.257 | 0.000 | −0.408 | 0.000 | −0.297 | 0.000 | −0.107 | 0.045 |
| | Quzhou | −0.323 | 0.000 | −0.396 | 0.000 | −0.151 | 0.000 | 0.433 | 0.000 |
| | Lishui | −0.223 | 0.000 | −0.495 | 0.000 | −0.223 | 0.000 | −0.014 | 0.436 |
| | Hefei | −0.138 | 0.093 | −0.352 | 0.000 | −0.326 | 0.000 | −0.100 | 0.021 |
| | Fuzhou | −0.233 | 0.000 | −0.342 | 0.000 | −0.183 | 0.000 | 0.289 | 0.000 |
| | Xiamen | −0.158 | 0.000 | −0.313 | 0.000 | −0.148 | 0.000 | 0.266 | 0.000 |
| | Nanchang | −0.393 | 0.000 | −0.251 | 0.000 | −0.290 | 0.000 | 0.338 | 0.000 |
| | Jinan | 0.211 | 0.000 | −0.219 | 0.000 | −0.285 | 0.000 | 0.124 | 0.000 |
| | Qingdao | −0.129 | 0.183 | −0.293 | 0.000 | −0.064 | 0.037 | 0.205 | 0.000 |

# References

Barnaba, F., Gobbi, G.P., 2004. Aerosol seasonal variability over the Mediterranean region and relative impact of maritime, continental and Saharan dust particles over the basin from MODIS data in the year 2001. Atmos. Chem. Phys. 4(9/10), 2367–2391.

Cao, J.J., Shen, Z.X., Chow, J.C., Watson, J.G., Lee, S.C., Tie, X.X., Ho, K.F., Wang, G.H., Han, Y. M., 2012. Winter and summer $PM_{2.5}$ chemical compositions in fourteen Chinese cities. J. Air Waste Manage. 62(10), 1214–1226.

Cao, C., Lee, X., Liu, S., Schultz, N., Xiao, W., Zhang, M., Zhao, L., 2016. Urban heat islands in China enhanced by haze pollution. Nat. Commun. 7, 12509.

Cai, W., Li, K., Liao, H., Wang, H., Wu, L., 2017. Weather conditions conducive to Beijing severe haze more frequent under climate change. Nature Clim. Change 7(4), 257–262.

Chang W, Zhan J., 2017. The association of weather patterns with haze episodes: Recognition by $PM_{2.5}$ oriented circulation classification applied in Xiamen, Southeastern China. Atmos. Res. 197, 425-436.

Chen, T., He, J., Lu, X., She, J., Guan, Z., 2016. Spatial and Temporal Variations of $PM_{2.5}$ and Its Relation to Meteorological Factors in the Urban Area of Nanjing, China. Int. J. Environ. Res. Public Health 13(9), 921.

Chen, Z., Cai, J., Gao, B., Xu, B., Dai, S., He, B., Xie, X., 2017. Detecting the causality influence of individual meteorological factors on local $PM_{2.5}$ concentration in the Jing-Jin-Ji region. Sci. Rep. 7, 40735

Chu, Y., Liu, Y., Li, X., Liu, Z., Lu, H., Lu, Y., Mao, Z., Chen, X., Li, N., Ren, M., Liu, F., Tian, L., Zhu, Z., Xiang, H., 2016. A Review on Predicting Ground $PM_{2.5}$ Concentration Using Satellite Aerosol Optical Depth. Atmosphere 7(10), 129.

Dai, W., Gao, J., Cao, G., & Ouyang, F., 2013. Chemical composition and source identification of $PM_{2.5}$ in the suburb of Shenzhen, China. Atmos. Res. 122, 391-400.

Daly, C., Gibson, W.P., Taylor, G.H., Johnson, G.L., Pasteris, P., 2002. A knowledge-based approach to the statistical mapping of climate. Clim. Res. 22(2), 99–113.

Dan, M., Zhuang, G., Li, X., Tao, H., Zhuang, Y., 2004. The characteristics of carbonaceous species and their sources in $PM_{2.5}$ in Beijing. Atmos. Environ. 38(21), 3443-3452.

Dawson, J.P., Adams, P.J., Pandis, S.N., 2007. Sensitivity of $PM_{2.5}$ to climate in the Eastern US: a modeling case study. Atmos. Chem. Phys. 7(16), 4295–4309.

Geng, G., Zhang, Q., Martin, R.V., van Donkelaar, A., Huo, H., Che, H., Lin, J., He, K., 2015.

Estimating long-term PM$_{2.5}$ concentrations in China using satellite-based aerosol optical depth and a chemical transport model. Remote Sens. Environ. 166, 262–270.

Guo, Y., Zeng, H., Zheng, R., Li, S., Barnett, A. G., Zhang, S., Zou, X., Huxley, R., Chen, W., Williams, G., 2016. The association between lung cancer incidence and ambient air pollution in China: a spatiotemporal analysis. Environ. Res. 144, 60–65.

Hu, D., Qiao, L., Chen, J., Ye, X., Yang, X., Cheng, T., Fang, W., 2010. Hygroscopicity of inorganic aerosols: size and relative humidity effects on the growth factor. Aerosol Air Qual. Res. 10, 255–264.

Hu, X., Waller, L.A., Al-Hamdan, M.Z., Crosson, W.L., Estes, M.G., Estes, S.M., Quattrochi, D.A., Sarnat, J.A., Liu, Y., 2013. Estimating ground-level PM$_{2.5}$ concentrations in the southeastern US using geographically weighted regression. Environ. Res. 121, 1–10.

Huang, F., Li, X., Wang, C., Xu, Q., Wang, W., Luo, Y., Tao, L., Gao, Q., Guo, J., Chen, S., Cao, K., Liu, L., Gao, Ni, Liu, X., Yang, K., Yan, A., Guo, X., 2015. PM$_{2.5}$ spatiotemporal variations and the relationship with meteorological factors during 2013-2014 in Beijing, China. PloS ONE. 10(11), e0141642.

Jones, A.M., Harrison, R.M., Baker, J., 2010. The wind speed dependence of the concentrations of airborne particulate matter and NO$_x$. Atmos. Environ. 44(13), 1682–1690.

Krivoruchko, K., 2012. Empirical Bayesian kriging. ESRI: Redlands, CA, USA. Available online at: http://www. esri. com/news/arcuser/1012/empirical-byesian-kriging.html

Lai, S., Zhao, Y., Ding, A., Zhang, Y., Song, T., Zheng, J., Song, T., Zheng, J., Lee, S., Zhong, L., 2016. Characterization of PM$_{2.5}$ and the major chemical components during a 1-year campaign in rural Guangzhou, southern China. Atmos. Res. 167, 208-215.

Li, G., Fang, C., Wang, S., Sun, S., 2016. The effect of economic growth, urbanization, and industrialization on fine particulate matter (PM$_{2.5}$) concentrations in China. Environ. Sci. Technol. 50(21), 11452–11459.

Li, J., Wang, G., Wang, X., Cao, J., Sun, T., Cheng, C., Meng, J., Hu, T., Liu, S., 2013. Abundance, composition and source of atmospheric PM$_{2.5}$ at a remote site in the Tibetan Plateau, China. Tellus B: Chem. Phys. Meteorol. 65(1), 20281.

Li, T., Shen, H., Zeng, C., Yuan, Q., Zhang, L., 2017. Point-surface fusion of station measurements and satellite observations for mapping PM$_{2.5}$ distribution in China: Methods and assessment. Atmos. Environ. 152, 477–489.

Li, T., Shen, H., Yuan, Q., Zhang, X., Zhang, L., 2017. Estimating ground-level PM$_{2.5}$ by fusing satellite and station observations: A geo-intelligent deep learning approach. arXiv preprint

arXiv:1707.03558.

Li, Y., Chen, Q., Zhao, H., Wang, L., Tao, R., 2015. Variations in $PM_{10}$, $PM_{2.5}$ and $PM_{1.0}$ in an urban area of the Sichuan Basin and their relation to meteorological factors. Atmosphere 6(1), 150–163.

Liu, J., Rühland, K. M., Chen, J., Xu, Y., Chen, S., Chen, Q., Huang, W., Xu, Q., Chen, F., Smol, J. P., 2017. Aerosol-weakened summer monsoons decrease lake fertilization on the Chinese Loess Plateau. Nature Clim. Change 7(3), 190–194.

Liu, Y., Zhao, N., Vanos, J.K., Cao, G., 2017. Effects of synoptic weather on ground-level $PM_{2.5}$ concentrations in the United States. Atmos. Environ. 148, 297–305.

Lu, D., Xu, J., Yang, D., Zhao, J., 2017. Spatio-temporal variation and influence factors of $PM_{2.5}$ concentrations in China from 1998 to 2014. Atmos. Pollut. Res.

Lu, X., Lin, C., Li, Y., Yao, T., Fung, J. C., Lau, A. K., 2017. Assessment of health burden caused by particulate matter in southern China using high-resolution satellite observation. Environ. Int. 98, 160–170.

Luo, J., Du, P., Samat, A., Xia, J., Che, M., Xue, Z., 2017. Spatiotemporal Pattern of $PM_{2.5}$ Concentrations in Mainland China and Analysis of Its Influencing Factors using Geographically Weighted Regression. Sci. Rep. 7, 40607.

Ma, Z., Hu, X., Huang, L., Bi, J., Liu, Y., 2014. Estimating ground-level $PM_{2.5}$ in China using satellite remote sensing. Environ. Sci. Technol. 48(13), 7436–7444.

Ma, X., Wang, J., Yu, F., Jia, H., & Hu, Y., 2016. Can MODIS AOD be employed to derive $PM_{2.5}$ in Beijing-Tianjin-Hebei over China? Atmos. Res. 181, 250-256.

Tai, A.P., Mickley, L.J., Jacob, D.J., 2010. Correlations between fine particulate matter, $PM_{2.5}$ and meteorological variables in the United States: Implications for the sensitivity of $PM_{2.5}$ to climate change. Atmos. Environ. 44(32), 3976–3984.

Tran, H. N., Mölders, N., 2011. Investigations on meteorological conditions for elevated $PM_{2.5}$ in Fairbanks, Alaska. Atmos. Res. 99(1), 39-49.

Wang, B., Wu, R., Lau, K.M., 2001. Interannual variability of the Asian summer monsoon: Contrasts between the Indian and the western North Pacific–East Asian monsoons. J. Clim. 14(20), 4073–4090.

Wang, Y., Ying, Q., Hu, J., Zhang, H., 2014. Spatial and temporal variations of six criteria air pollutants in 31 provincial capital cities in China during 2013–2014. Environ. Int. 73, 413–422.

World Health Organization, UNAIDS., 2006. Air quality guidelines: global update 2005. World Health Organization.

Wu, J., Zhu, J., Li, W., Xu, D., Liu, J., 2017. Estimation of the $PM_{2.5}$ health effects in China during 2000–2011. Environ. Sci. Pollut. Res. 24(11), 10695–10707.

Wu, P., Ding, Y., Liu, Y., 2017. Atmospheric circulation and dynamic mechanism for persistent haze events in the Beijing-Tianjin-Hebei region. Advances in Atmos. Sci. 34(4), 429–440.

Xie, Y., Dai, H., Dong, H., Hanaoka, T., Masui, T., 2016. Economic impacts from $PM_{2.5}$ pollution-related health effects in China: A provincial-Level analysis. Environ. Sci. & Technol. 50(9), 4836–4843.

Xu, J., Yan, F., Xie, Y., Wang, F., Wu, J., Fu, Q., 2015. Impact of meteorological conditions on a nine-day particulate matter pollution event observed in December 2013, Shanghai, China. Particuology 20, 69–79.

Yan, S., Cao, H., Chen, Y., Wu, C., Hong, T., Fan, H., 2016. Spatial and temporal characteristics of air quality and air pollutants in 2013 in Beijing. Environ. Sci. Pollut. Res. 23(14), 13996–14007.

Ye, X., Chen, J., 2013. Haze and hygroscopic growth. Chin. J. Nat. 35(5), 337–341.

Zhang, B., Jiao, L., Xu, G., Zhao, S., Tang, X., Zhou, Y., Gong, C., 2017. Influences of wind and precipitation on different-sized particulate matter concentrations ($PM_{2.5}$, $PM_{10}$, $PM_{2.5-10}$). Meteorol. Atmos. Phys. 1–10.

Zhang, F., Cheng, H.R., Wang, Z.W., Lv, X.P., Zhu, Z.M., Zhang, G., Wang, X.M., 2014. Fine particles ($PM_{2.5}$) at a CAWNET background site in Central China: Chemical compositions, seasonal variations and regional pollution events. Atmos. Environ. 86, 193–202.

Zhang, H., Wang, Y., Hu, J., Ying, Q., Hu, X. M., 2015. Relationships between meteorological parameters and criteria air pollutants in three megacities in China. Environ. Res. 140, 242–254.

Zhang, H., Wang, Y., Park, T. W., & Deng, Y., 2017. Quantifying the relationship between extreme air pollution events and extreme weather events. Atmos. Res. 188, 64-79.

Zhang, H., Wang, Z., Zhang, W., 2016. Exploring spatiotemporal patterns of $PM_{2.5}$ in China based on ground-level observations for 190 cities. Environ. Pollut. 216, 559–567.

Zhao, C., Tie, X., Lin, Y., 2006. A possible positive feedback of reduction of precipitation and increase in aerosols over eastern central China. Geophys. Res. Lett. 33(11).

Zheng, M., Salmon, L.G, Schauer, J.J., Zeng, L., Kiang, C.S., Zhang, Y., Cass, G.R., 2005. Seasonal trends in $PM_{2.5}$ source contributions in Beijing, China. Atmos. Environ. 39(22),

3967–3976.